\documentstyle[aps,prd]{revtex}
\newcommand{\be}{\begin{equation}}
\newcommand{\ee}{\end{equation}}
\newcommand{\ba}{\begin{eqnarray}}
\newcommand{\ea}{\end{eqnarray}}

\begin{document}
\title{Yang--Mills fields as optical media} 
\author{R. Aldrovandi and A. L. Barbosa} 
\address{Laboratoire de  Gravitation et Cosmologie Relativistes \\
Universit{\'e} Pierre et Marie Curie,\  CNRS/ESA 7065  \\
Tour 22/12  4{\`e}me {\'e}tage  BP 142 \\ 
4 \ Place Jussieu \ Cedex 05 \\ 75252 Paris \ France} 
\date{\today}
\maketitle
\begin{abstract}
A geometrization of the Yang--Mills field, by which an SU(2) gauge 
	theory becomes equivalent to a 3-space geometry -- or optical 
	system -- is examined.  In a first step, ambient space remains 
	Euclidean and current problems on flat space can be looked at from 
	a new point of view.  The Wu--Yang ambiguity, for example, appears 
	related to the multiple possible torsions of distinct 
	metric--preserving connections.  In a second step, also the 
	ambient space becomes curved.  In general, the strictly 
	Riemannian, metric sector plays the role of an arbitrary host 
	space, with the gauge field represented by a contorsion.  For some 
	field configurations, however, it is possible to obtain a purely 
	metric representation.  In those cases, if the space is symmetric homogeneous the 
	Christoffel connections are automatically solutions of the Yang--Mills equations.	
\end{abstract}


	\section{Introduction}	 %

Our main intuitive guide to interactions is, ultimately, the 
nonrelativistic idea of potential.  We recall, for example, the 
phenomenological potentials used with reasonable success in low energy 
hadron spectroscopy: a Coulomb-like term, plus a linear potential 
providing for the confining behavior, are thought to represent the 
nonrelativistic limit of the time components $A^{a}{}_{0}$ of the 
gauge potential in chromodynamics.  The Wilson--loop criterion for 
confinement gives potentials of that kind in the nonrelativistic 
limit, and is thereby justified.  There are, however, great advantages 
in the use of the (temporal, or Weyl) gauge $A^{a}{}_{0} = 0$, which 
is obviously incompatible with such a view.  The merits of this gauge 
(Feynman, 1977) are particularly relevant when associated to the 
Hamiltonian formalism (Jackiw, 1980).  The major qualitative 
characteristics of gauge fields, such as shielding and confinement, 
are nowadays believed to be essentially non-perturbative, and the best 
approach available to consider global aspects is precisely the 
Hamiltonian formalism.  On the other hand, the impossibility of 
thinking in terms of a potential inhibits intuition.  It would be nice 
to have some other qualitative guide in its stead.  This note is 
intended to call attention to the possibility of -- at least in some 
cases -- using a Geometrical Optics analogy.  A static sourceless 
SU(2) gauge field configuration can, in the temporal gauge, be 
`'geometrized'' to become equivalent to a metric plus a torsion on 
3-space.  A metric on a 3-space is a simple optical system (Guillemin 
and Sternberg, 1977), as it can be seen as the dielectric tensor 
$\epsilon_{i j}$ of a medium to which torsion will add some defects 
(Aldrovandi and Pereira, 1995).  An optical picture, with refractive 
indices and defects taking the place of potentials, could be an 
alternative source of ideas.
	
Geometrization of the SU(2) theory was proposed by Lunev (Lunev, 
1992), and by Freedman, Haagensen, Johnson and Latorre in a tentative 
to arrive at a description of gauge fields in terms of invariants.  In 
particular, it was part of a program (Freedman, 1993) 
to solve a great problem in the quantization of the Hamiltonian scheme 
--- the implementation of Gauss' law.  In that pursuit some 
assumptions were made which were not necessary to the simpler aim of 
establishing a geometrized version of the theory.  We present in the 
following a minimal approach, using only the hypotheses strictly 
necessary to that particular end.  It turns out that, in the generic 
case, the metric sector is highly arbitrary and acts as a ``host'' 
space, on which the ``guest'' gauge potential is represented by the 
contorsion tensor.  For some field configurations, however, it is 
possible to choose a metric which alone contains all the information.
	
We begin by recalling the main aspects of the Hamiltonian approach to 
Yang--Mills theory, in which time and Euclidean 3-space ${\bf E}^{3}$ 
are clearly separated.  We show then how to transcribe the field 
equations into those of a geometry on ${\bf R}^{3}$.  Complete 
geometrization would lead immediately to gauge theories on curved 
spaces.  We proceed consequently in two steps.  In the first, only the 
indices related to the Lie algebra are ``geometrized'', while ambient 
space remains the flat space ${\bf E}^{3}$.  The main geometrical 
ideas are already present, but we remain able to discuss questions 
turning up in flat space.  Some of them are seen under a new angle.  
For example, the Wu--Yang ambiguity is related to the multiplicity of 
torsion tensors with a fixed curvature.  In the second step, also the 
ambient ${\bf R}^{3}$ is endowed with a new, rather arbitrary metric.  
The Yang--Mills equations appear then written on curved spaces.  If 
such spaces are torsionless homogeneous symmetric spaces, their very 
Christoffel connections are solutions.  In such cases, the full 
geometrization exhibits field configurations which are completely 
equivalent to simple optical systems.  The group really considered is 
SU(2), for which the geometry unfolds itself in a quite natural way.  
In our notation, greek indices run from 0 to 3 and latin indices from 
1 to 3.

	\section{Hamiltonian formalism}	  %

In the Hamiltonian approach (Faddeev and Slavnov, 1978; Itzykson and 
Zuber, 1980; Ramond, 1981) to the Yang-Mills equations, the canonical 
coordinates are the vector potential components $A^{a}{}_{k}$ and, 
once the Lagrangian ${\mathcal L} = - \frac{1}{4}F_{a}{}^{\mu \nu} 
F^{a}{}_{\mu \nu}$ is given, the conjugate momenta $\Pi^{a}{}_{k}$ are 
the electric fields:
	
\be \Pi^{a i} = \frac{\delta {\mathcal L}}{\delta 
\partial_{0}A^{a}{}_{i}} = F^{a i 0} = E^{a i} = \partial^{i}A^{a 0} - 
\partial^{0} A^{a i} + f^{a}{}_{bc}A^{b i}A^{c 0} \ \; .  \ee The 
action can then be rewritten in the form
	
\be S = 2 \int d^{4}x \ tr \left[ \partial^{0} {\bf A} \cdot {\bf E} + 
{\textstyle \frac{1}{2}} ({\bf E}^{2} + {\bf B}^{2}) - A_{a}{}{}^{0} 
G^{a}(x) \right] , \label{action} \ee where
	
\be G^{a}(x) = D_{k} E^{a k} = \partial_{k}E^{a k} + 
f^{a}{}_{bc}A^{b}{}_{k}E^{c k} \;
	\label{GaussEspr} 
\ee (we have profited to define the derivative $D_{k}$).  A 
constraint, a redefinition of terms and two dynamic equations come 
out.  The first two are
\begin{enumerate}

\item the Gauss law, which states the vanishing of (\ref{GaussEspr}):
	
\be G^{a}(x) = \partial_{k}E^{a k} + f^{a}{}_{bc}A^{b}{}_{k} E^{c k} = 
0 \label{GaussLaw} \ee(where we see that $A_{0}$ is, in action 
(\ref{action}), a Lagrange multiplier enforcing Gauss' law);

\item the expression of the magnetic field in terms of $A^{a}{}_{k}$,

\be B^{a i} = {\textstyle \frac{1}{2}} \epsilon^{i j k} F^{a}{}_{j k} 
= \epsilon^{i j k} \left(\partial_{j}A^{a}{}_{k} + {\textstyle 
\frac{1}{2}} f^{a}{}_{bc}A^{b}{}_{j} A^{c}{}_{k} \right) \; .
\label{Bexpr}
\ee The dynamic equations are Hamilton's equations:

\item the time variation of the vector potential,

\be \frac{1}{c} \frac{\partial}{\partial t} A^{a}{}_{i} = - 
E^{a}{}_{i} - \partial_{i} A^{a 0} + f^{a}{}_{bc}A^{b 0}{} A^{c}{}_{i} 
\; ; \ee

\item Amp\`ere's law, here in the role of the force law:

\be \frac{1}{c} \frac{\partial}{\partial t} E^{a i} = ({\mathbf 
\nabla} \times {\bf B})^{a i} + \epsilon^{i}{}_{jk} f^{a}{}_{bc}A^{b 
j} B^{c k} + f^{a}{}_{bc}A^{b 0} E^{c i} \; .  \ee Consider now the 
gauge $A^{a}{}_{0} = 0$.  The Hamiltonian is

\be H = {\textstyle \frac{1}{2}} \int d^{3}x \ tr \left[ ({\bf E}^{2} 
+ {\bf B}^{2}) \right] \; .  \label{Hamiltonian} \ee
\end{enumerate}
A static $A^{a}{}_{i}$ leads to $E^{a}{}_{i} = 0$.  Gauss' law is 
automatically satisfied and Amp\`ere's law reduces to

\be ({\mathbf \nabla} \times {\bf B})^{a i} + \epsilon^{i}{}_{jk} 
f^{a}{}_{bc}A^{b j} B^{c k} = 0 \; .  \label{AmpLaw} \ee

Notice that (\ref{Bexpr}) fixes B once A is given, but not vice-versa.  
This is the Wu-Yang ambiguity (Wu and Yang, 1975): many inequivalent 
gauge fields $A^{a}{}_{i}$ can correspond to the same magnetic field 
$B^{a}{}_{i}$.  A consequence is that the $B^{a}{}_{i}$'s cannot be 
used as coordinates.

	\section{Stepping into Geometry}	  %

Let us stress beforehand that many different metrics can be defined on 
the same space (for us, ``space'' will mean only a differentiable 
manifold).  The best examples of such a metric multiplicity are 
provided precisely by optical systems (Luneburg, 1966), whose 
treatment is greatly eased by the simultaneous use of the Euclidean 
metric $\delta_{ij}$ of ${\bf E}^{3}$ and of the dielectric tensor 
$\epsilon_{ij}$.  Isotropic media have $\epsilon_{ij}$ = $n^{2} 
\delta_{ij}$, with $n$ the refractive index and correspond to 
conformally--flat 3-spaces.  Notice that by ${\bf E}^{3}$ we 
understand the usual Euclidean metric space, the space ${\bf R}^{3}$ 
of real ordered triples endowed with its unique differentiable 
structure and with the additional proviso that length measurements are 
performed supposing that $ds^{2}$ = $\delta_{ij} dx^{i} dx^{j}$ = 
$dx^{2} + dy^{2} + dz^{2}$.  An optical system will be the same 
differentiable manifold ${\bf R}^{3}$ but with optical lengths 
instead, measured with the dielectric metric $dl^{2} = \epsilon_{ij} 
dx^{i} dx^{j}$.  Another example is given by the group SU(2), whose 
manifold is the 3-sphere $S^{3}$.  It has the ``natural'' spherical 
metric which comes up when $S^{3}$ is seen as an imbedded submanifold 
of the Euclidean space ${\bf E}^{4}$, but it has also the flat 
Killing-Cartan metric $\gamma_{a b} = \delta_{a b}$, which is more 
important from the algebraic point of view. 
		
Now, looking back to what comes out in the Hamiltonian formalism: The 
only equations remaining in the static case are (\ref{Bexpr}) and 
(\ref{AmpLaw}).  They can be easily rewritten in terms of a spatial 
geometry in the following way.  First, we notice that both the spaces 
involved are 3-dimensional, on which vectors are equivalent to 
antisymmetric 2-tensors.  Indices can be trivially ``dualized''.  We 
can redefine the gauge potential as the connection

\be \omega^{a}{}_{c k} = 
\epsilon^{a}{}_{bc} \  A^{b}{}_{k}   \label{omegaA}
\ee 
with curvature

\be R^{a}{}_{b i j} = \epsilon^{a}{}_{c b} \ F^{c}{}_{i j} = 
\epsilon_{i j k} \ \epsilon^{a}{}_{c b} \ B^{c k},
\label{BintoR} 
\ee
in terms of which (\ref{Bexpr}) becomes 

\be R^{a}{}_{b i j} = \partial_{i} \ \omega^{a}{}_{b j} - \partial_{j} 
\ \omega^{a}{}_{b i} + \omega^{a}{}_{c i} \ \omega^{c}{}_{b j} - 
\omega^{a}{}_{c j} \ \omega^{c}{}_{b i} \; .  \ee We find immediately 
that, for any object of index--type $W^{a}$,

\be [D_{i}, D_{j}] \ W^{a} = R^{a}{}_{b i j}\ W^{b} \; .  
\label{commut1} \ee

Second, we notice that the flat sphere SU(2), with the metric 
$\gamma_{a b}$, is isomorphic to ${\bf E}^{3}$ and, consequently, to 
the space tangent to the ambient space on which the differential 
equations are written.  There are actually infinite such isomorphisms, 
each one realized by a dreibein field $h^{a}{}_{i}$ (we shall be using 
the letters a, b, c \ldots as isotopic spin indices and i, j, r, s, 
\ldots as ambient space indices).  A dreibein field, together with 
its inverse $h_{b}{}^{j}$, can be used to change tensor indices as in

\be R^{r}{}_{s i j} = h^{r}{}_{a} \ h^{b}{}_{s} \ R^{a}{}_{b i j} \; .
\label{firstR}  
\ee Connections, however, are not truly tensorial.  Only the last (in 
our notation), derivative index is a covector index.  The other are 
not, and are translated according to

\be \Gamma^{i}{}_{j k} = h^{i}{}_{a} \ \omega^{a}{}_{b k} \ 
h^{b}{}_{j} + h^{i}{}_{c} \ \partial_{k} \ h^{c}{}_{j} .  
\label{Gamomega} \ee This comes from the requirement that the 
covariant derivative remain covariant under change of basis.  The 
connection $\omega^{a}{}_{b k}$ would represent, in the absence of 
torsion, the Ricci rotation coefficients (Chandrasekhar 1992) or, if 
we borrow from the usual treatment of the Dirac equation on curved 
spaces, the spin connection.  With the above transformations, the 
equations (\ref{Bexpr}) and (\ref{AmpLaw}) become
 
\be R^{r}{}_{s i j} = \partial_{i} \ \Gamma^{r}{}_{s j} - \partial_{j} 
\ \Gamma^{r}{}_{s i} + \Gamma^{r}{}_{k i} \ \Gamma^{k}{}_{s j} - 
\Gamma^{r}{}_{k j} \ \Gamma^{k}{}_{s i} \; \label{curvature1} \ee %
and %
\be \partial_{j} \ R^{r}{}_{s i}{}^{j} + \Gamma^{r}{}_{k}{}_{j} \ 
R^{k}{}_{s i}{}^{j} - \Gamma^{k}{}_{s}{}_{j} \ R^{r}{}_{k i }{}^{j} = 
0 \; , \label{Ampere} \ee %
where now all the indices refer to the ambient space.  Notice that the 
dreibeine are quite arbitrary.  Equation (\ref{curvature1}) simply 
defines $R^{r}{}_{s i j}$ as the curvature of the connection 
$\Gamma^{r}{}_{s i}$, but the nine equations (\ref{Ampere}), stating 
Amp\`ere's law, keep their dynamical role.

Another characteristic of the connection, its torsion, will be 
given by 

\be T^{a}{}_{i j} = \partial_{i} \ h^{a}{}_{j} - \partial_{j} \ 
h^{a}{}_{i} + \omega^{a}{}_{c i} \ h^{c}{}_{j} - \omega^{a}{}_{c j} \ 
h^{c}{}_{i}, \ee or, after transmuting the indices, \be T^{k}{}_{i j} 
= \; - \; \Gamma^{k}{}_{[i j]} \; .  \ee %
We are introducing the notation [ij] for antisymmetrized indices 
without any numerical factors, and we shall use (ij) for 
symmetrization.  This will lead, for example, to the identity

\be \Gamma^{k}{}_{ij} = {\textstyle \frac{1}{2}} 
\left(\Gamma^{k}{}_{(ij)} + \Gamma^{k}{}_{[ij]} \right) \; .  
\label{decomp1} \ee

Some formal expressions are of interest to ease manipulations: first,

\be \Gamma^{k}{}_{i j} = h_{a}{}^{k} D_{j} h^{a}{}_{i} \; \ee and its 
consequence \be T^{k}{}_{i j} = h_{a}{}^{k} D_{[i} h^{a}{}_{j]} \; .  
\ee Equation (\ref{commut1}) implies %
\be R^{r}{}_{s i j} = h_{a}{}^{r} [D_{i}, D_{j}] h^{a}{}_{s} \; .  
\label{commut2} \ee Finally, the Bianchi identity for torsion, \be 
D_{[i} T^{a}{}_{j k]} = R^{a}{}_{[i j k]} \; .  \label{BianchiFirst} 
\ee

The dreibein field will define a metric on ${\bf R}^{3}$ by \be g_{ij} 
= \gamma_{a b} h^{a}{}_{i} h^{b}{}_{j} .  \ee This metric is automatically preserved by $\Gamma^{k}{}_{ij}$.  
Indeed, the metric compatibility condition (which means that the 
metric is parallel-transported by $\Gamma$)

\be \partial_{k} g_{ij} = \Gamma_{ijk} + \Gamma_{jik} = \Gamma_{(ij)k} 
\label{compatibility} \ee is a simple consequence of (\ref{Gamomega}).

Thus, in the transcription of Yang--Mills fields into a spatial 
geometry induced by a dreiben field, the gauge potential is transmuted 
into a connection which automatically preserves the metric defined by 
the dreibeine.  Given a dreibein field and the original $A$, the 
connection $\Gamma$ is unique.  This is due to the Ricci lemma (Greub, 
1972), which reads: given a metric $g$ and any tensor of type 
$T^{k}{}_{i j}$, there is one and only one connection which preserves 
$g$ and has torsion equal to $T^{k}{}_{i j}$.
 
Both curvature and torsion are properties of a connection (Kobayashi 
and Nomizu l963).  There are in principle an infinity of connections 
which preserve a given metric $g_{ij}$.  Of all these connections only 
one, the Levi-Civita connection ${\stackrel{\circ}{\Gamma}}$, has 
vanishing torsion (a weak version of the Ricci lemma).  The others 
differ from that privileged one precisely by their torsions.  The 
components of the Levi-Civita connection are the well-known 
Christoffel symbols

\be {{\stackrel{\circ}{\Gamma}}}^{k}{}_{i j} = {\textstyle 
\frac{1}{2}} g^{k r} \left[\partial_{i} g_{j r} + \partial_{j} g_{i r} 
- \partial_{r} g_{i j} \right] \; \label{Christoffel} .  \ee The 
strictly Riemannian curvature ${\stackrel{\circ}{R}}$ will be
  
\be {\stackrel{\circ}{R}}^{r}{}_{s i j} = \partial_{i} 
{\stackrel{\circ}{ \ \Gamma}}^{r}{}_{s j} - \partial_{j} 
{\stackrel{\circ}{\Gamma}}^{r}{}_{s i} + 
{\stackrel{\circ}{\Gamma}}^{r}{}_{k i} 
{\stackrel{\circ}{\Gamma}}^{k}{}_{s j} - 
{\stackrel{\circ}{\Gamma}}^{r}{}_{k j} 
{\stackrel{\circ}{\Gamma}}^{k}{}_{s i} \; .  \label{curvatureball} \ee

A connection exhibits torsion in the generic case.  Now, given a 
general connection $\Gamma^{k}{}_{i j}$ preserving a metric $g_{ij}$, 
it can always be written in the form

\be \Gamma^{k}{}_{i j} = {\stackrel{\circ}{\Gamma}}^{k}{}_{i j} - 
K^{k}{}_{i j}, \label{decomp2} \ee %
where $K^{k}{}_{i j}$ is its contorsion tensor.  Any two connections 
differ by some tensor, but here metric compatibility gives an extra 
constraint: contorsion is fixed by the torsion tensor,

\be K^{k}{}_{i j} = {\textstyle \frac{1}{2}} \left[T^{k}{}_{i j} + 
T_{i j}{}^{k} + T_{j i}{}^{k}\right] \; . \label{contorsion} \ee 

This comes from the comparison of two expressions for 
${\stackrel{\circ}{\Gamma}}$: one obtained by substituting 
(\ref{compatibility}) in (\ref{Christoffel}) three times; the other by 
using (\ref{decomp1}) in (\ref{decomp2}).  As both T and K are 
tensors, this relationship holds in any basis.  Notice that the 
decompositions (\ref{decomp1}) and (\ref{decomp2}) are not the same.  
The two last terms in (\ref{contorsion}) show a symmetric contribution 
of contorsion to $\Gamma$: $K^{k}{}_{(i j)}$ = $T_{(i j)}{}^{k}$.  In 
consequence, \be \Gamma^{k}{}_{(i j)} = 
{\stackrel{\circ}{\Gamma}}^{k}{}_{i j} - T_{(i j)}{}^{k}\ee and \be 
\Gamma^{k}{}_{[i j]} = \; - \; K^{k}{}_{[i j]} = \; - \; T^{k}{}_{i 
j}.  \ee The property \be K_{(ki) j} = 0 \; \ee follows from the fact 
that ${\stackrel{\circ}{\Gamma}}$ satisfies (\ref{compatibility}) 
independently.

The presence of torsion changes curvature.  Indeed, the total 
curvature (\ref{curvature1}) is

\be R^{r}{}_{s i j} = {{\stackrel{\circ}{R}}}^{r}{}_{s i j} - 
M^{r}{}_{s i j} \; , \label{curvature2} \ee
where

\[ M^{r}{}_{s i j} = \partial_{i} K^{r}{}_{s j} - \partial_{j} 
K^{r}{}_{s i} + {\stackrel{\circ}{\Gamma}}^{r}{}_{n i} K^{n}{}_{s j} + 
K^{r}{}_{n i} {\stackrel{\circ}{\Gamma}}^{n}{}_{s j} - 
{\stackrel{\circ}{\Gamma}}^{r}{}_{n j} K^{n}{}_{s i} - K^{r}{}_{n j} 
{\stackrel{\circ}{\Gamma}}^{n}{}_{s i}
\]
\be \; \; \; \; \; \; \; \; \; \; \; \; \; \; \; - K^{r}{}_{n i} 
K^{n}{}_{s j} + K^{r}{}_{n j} K^{n}{}_{s i} \; .  \label{curvtors} \ee

We have thus obtained a geometrized version of the Yang--Mills system 
in Euclidean space.  The original SU(2) connection $A$ has been 
transformed into a linear connection on ${\bf E}^{3}$, which will be 
``felt'' by SU(2) non-singlet particles.  Notice that, once a 
particular dreibein field is used, $g_{ij}$, 
${\stackrel{\circ}{\Gamma}}$ and ${\stackrel{\circ}{R}}^{r}{}_{s i j}$ 
are fixed.  The metric being arbitrary, the trivial choice would be a 
flat host space: ${\stackrel{\circ}{\Gamma}} = 0$, 
${\stackrel{\circ}{R}} = 0$.  In that case, the equations reduce to
\[
M^{r}{}_{s i j} = \partial_{i} K^{r}{}_{s j} - \partial_{j} K^{r}{}_{s 
i} - K^{r}{}_{k i}K^{k}{}_{s j} - K^{r}{}_{k j}K^{k}{}_{s i} \; ;
\]
\[
\partial_{j} M^{r}{}_{s i}{}^ {j} - K^{r}{}_{k j} M^{k}{}_{s i}{}^ {j} 
+ K^{k}{}_{s j} M^{r}_{k i}{}^ {j} = 0 \; .
\]

These are just the equations (\ref{Bexpr}) and (\ref{AmpLaw}) we 
started from, with the gauge potential transmuted into a contorsion by 
trivial dreibeine.  Only the indices related to the gauge Lie algebra 
have been changed into 3-space indices up to now.  And, as we treat 
the new indices (r, s, \ldots ) on an equal footing with the original, 
holonomic ambient 3-space indices (i, j, \ldots ), what we are 
actually doing is to choose a new basis for the algebra at each point 
of the ambient space (in the language of the sixties, we are 
geometrizing the ``internal space'').  This is better understood if we 
consider the complete expressions of the algebra--valued differential 
forms involved.  In (\ref{firstR}), for example, what we have is
\begin{eqnarray*}
R = {\textstyle \frac{1}{2}} \ J_{c}{}{}^{b} \ R^{c}{}_{b i j} dx^{i} 
\wedge dx^{j} = {\textstyle \frac{1}{2}} \ J_{c}{}{}^{b} \ h^{c}{}_{r} 
h^{s}{}_{b} R^{r}{}_{s i j} dx^{i} \wedge dx^{j} \\
\; \;  \; \; \; \; \;  \; \;  \; \; \; \; \;   \; \;  \; \; \; \; \;   \; \;  \; \; \; \; \;  
=  {\textstyle \frac{1}{2}} \ J_{r}{}{}^{s} (x) R^{r}{}_{s i j} dx^{i} 
\wedge dx^{j} \; .
\end{eqnarray*}
We remain, for the time being, on the original flat space.  We shall 
later discuss the meaning of ``geometrizing'' the ambient space 
indices as well.
 
A few words on the Wu--Yang ambiguity.  Two distinct gauge potentials which 
have the same curvature are called ``copies''.  It should be said, to 
begin with, that there are no copies in the relationship between 
${\stackrel{\circ}{\Gamma}}$ and ${\stackrel{\circ}{R}}$.  There 
exists always around each point $p$ a system of coordinates in which 
${\stackrel{\circ}{\Gamma}}$ = 0 at $p$, so that the usual expression 
$R = d {\stackrel{\circ}{\Gamma}}$ + ${\stackrel{\circ}{\Gamma}}$ 
${\stackrel{\circ}{\Gamma}}$ reduces to ${\stackrel{\circ}{R}}$ = $d 
{\stackrel{\circ}{\Gamma}}$, which can be integrated to give locally 
${\stackrel{\circ}{\Gamma}}$ in terms of ${\stackrel{\circ}{R}}$.  
This is analogous to the equivalence principle which protects standard 
General Relativity from ambiguity, but holds only for symmetric, 
torsionless connections.  On the other hand, general linear 
connections exhibit copies in a natural way, as they can, in 
principle, have the same curvature and different torsions.  For 
example, each solution $K^{r}{}_{s i}$ (if any) of $M^{r}{}_{s i j} = 
0$ in (\ref{curvtors}) will lead to a copy of the Levi-Civita 
connection.  Instead of solving this equation, however, it is simpler to 
take the difference between the two Bianchi identities, which leads to an 
algebraic condition for the non-existence of copies 
(Roskies 1977; Calvo 1977).  Though powerful general results have 
been found on the problem (Mostow, 1980; Doria 1981), there seems to 
be no simple, systematic, calculating view of the problem.

Suppose now we start with two distinct potentials $A$ and $A'$ and 
transcribe them into 3-space geometries using the same dreibein field.  
As $g$ and ${\stackrel{\circ}{\Gamma}}$ will be the same, 
their difference will be in their contorsions.  
This will lead to different curvatures and torsions.  If, however, $A$ 
and $A'$ have the same curvature, only torsion will remain to 
distinguish them.  Thus, copies are 
``classified'' by torsions.  This is trivial for linear connections, 
but the good thing of the geometrization given above is exactly that: 
we can transfer to gauge potentials, which are connections related to 
internal groups, some of the properties of the linear, external 
connections.

	Notice also that discussions on the ambiguity are not, in general, 
	concerned with solutions and mostly ignore the dynamic classical 
	equations.  Non-solutions are very important because 
	they appear as off-shell contributions in the quantum case. The 
	geometric formulation has been used to produce examples of 
	continuous sets of copies (Freedman  and Khuri 1994).
	
We study now cases in which it is possible to choose the metric so as 
to completely absorb the gauge field.

	\section{Isotropic  Optics}	  %

Each choice of dreibeine will provide a different transcription into a 
3-space geometry.  A natural question is whether it is possible to 
choose them so as to absorb the gauge field entirely in the metric 
sector alone, dispensing with the torsion field.  Is it possible to 
transmute the gauge field into pure Optics ?  This would mean finding 
a dreibein field inducing a metric whose Levi-Civita connection 
${\stackrel{\circ}{\Gamma}}$ coincides with the transcript $\Gamma$ of 
$A^{a}{}_{j}$.

Consider dreibeine of the form

\begin{equation}
h_{\hspace{0.1cm}i}^{a}=\delta 
_{\hspace{0.1cm}i}^{a}\hspace{0.1cm}f(r) \; ,
\end{equation}
where $f(r)$ is any function depending only on the distance $r$ to 
some fixed origin.  The metric they define,

\begin{equation}
g_{ij}=\delta _{ij}\left[ \hspace{0.1cm}f(r)\right] ^{2} \; ,
\end{equation}
has the Levi-Civita connection

\begin{equation}
{\stackrel{\circ }{\Gamma }}_{\hspace{0.1cm}ij}^{k}= (\delta 
_{\hspace{0.1cm}j}^{k}x_{i}+\delta _{\hspace{0.1cm}%
i}^{k}x_{j}-\delta _{ij}x^{k}) \; \frac{1}{rf}\frac{\partial f%
}{\partial r}  \; ,  \label {a}
\end{equation}
with curvature
\[
\stackrel{\circ }{R}_{\hspace{0.1cm}sij}^{t}=\frac{1}{f}\frac{\partial f}{%
\partial r}\left( \frac{2}{r}+\frac{1}{f}\frac{\partial f}{\partial 
r}%
\right) (\delta _{\hspace{0.1cm}j}^{t}\delta _{si}-\delta _{\hspace{0.1cm}%
i}^{t}\delta _{sj})   \; \; +  \]
\begin{equation}
 \; \;  \; \frac{1}{r^{2}f}\left[ \frac{\partial 
^{2}f}{\partial r^{2}}-\frac{1}{r}%
\frac{\partial f}{\partial r}-\frac{2}{f}\left( \frac{\partial f}{\partial r}%
\right) ^{2}\right] (\delta _{\hspace{0.1cm}j}^{t}x_{i}x_{s}-\delta
_{sj}x^{t}x_{i}-\delta _{\hspace{0.1cm}i}^{t}x_{j}x_{s}+\delta
_{si}x^{t}x_{j}).
\label {b}
\end{equation}

Looking for solutions of the Yang-Mills equations, we take (\ref{a}) 
and (\ref{b}) into (\ref{Ampere}) and find
\begin{equation}
\frac{\partial ^{3}f}{\partial r^{3}}+\frac{1}{rf}\left( \frac{\partial f}{%
\partial r}\right) ^{2}-\frac{5}{f}\frac{\partial f}{\partial r}\frac{%
\partial ^{2}f}{\partial r^{2}}+\frac{5}{f^{2}}\left( \frac{\partial f}{%
\partial r}\right) ^{3}=0     \; .     \label{c}
\end{equation}

Using equations (\ref{Gamomega}) and (\ref{omegaA}) in $(\ref{a})$ we 
find the potential
\begin{equation}
A_{\hspace{0.1cm}j}^{d}=-\frac{1}{rf}\frac{\partial f}{\partial r}\epsilon _{%
\hspace{0.1cm}jk}^{d}x^{k} \; .  \label{d}
\end{equation}
From (\ref{b}), (\ref{firstR}) and (\ref{BintoR}) we have the magnetic field
\[
B_{\hspace{0.1cm}j}^{d}=\delta _{\hspace{0.1cm}j}^{d}\left[ \frac{2}{rf}%
\frac{\partial f}{\partial r}+\frac{4}{f^{2}}\left( \frac{\partial f}{%
\partial r}\right) ^{2}-\frac{1}{f}\frac{\partial ^{2}f}{\partial r^{2}}%
\right]  \]
\begin{equation}
\; \; \; \; \; \; \; \; \; \; \; \; \; \; \; \; \; \; \; \; \; \; \; 
\; \; \; \; \; \; \; \; \; \; \; \; \; \; \; \; \; + \; \; 
\frac{1}{r^{2}f}\left[ \frac{\partial ^{2}f}{\partial 
r^{2}}-\frac{1}{r}%
\frac{\partial f}{\partial r}-\frac{2}{f}\left( \frac{\partial f}{\partial r}%
\right) ^{2}\right] x_{j}x^{d}
 \label{e}
\end{equation}

Any solution of (\ref{c}) will lead to a solution of (\ref{AmpLaw}) 
given by (\ref{d}) and (\ref{e}).

Let us discuss a few particular cases.  Consider 
$\hspace{0.1cm}f(r)=\frac{1}{r^{q}}$, defined in all points of space 
except $r=0$.  The result of introducing it into (\ref{c}) is an 
equation for $q$,
\begin{equation}
q(2-q)(1-q)=0 \;  , \label{f}
\end{equation}
which three obvious solutions.  Cases $q=0$ and $q=2$ lead to trivial 
solutions: both the potential and the curvature are zero in the first 
case, and the second corresponds to a non-vanishing potential with 
zero curvature.  The only non-trivial solution is $q=1$, $\hspace{ 
0.1cm}f(r)=\frac{1}{r},$ the well-known Wu-Yang monopole (Wu and Yang 
1975)%
\ba A^{a}{}_{j} = \epsilon^{a}{}_{j k} \; \frac{x^{k}}{r^{2}} \; 
; \label{WYmonopA} \\
B^{a}{}_{j} =  \; -  \; \frac{x^{a} x_{j}}{r^{4}} \; .
\label{WYmonopB}
\ea

Expressions of the type $\exp (\pm r^{q}),\exp (1/(1-qr)),$ $\exp 
(1/(1-r^{q})),$ $\exp (\pm qr^{2})$ are only real solutions for $q = 
0$.  Expressions of the type $\left[ r/(1-qr)\right] $ and $\left[ 
r/(1-qr^{2}) \right] $ have only complex solutions for $q$.  The 
monopole is the only non-trivial solution found.  No torsion is 
necessary in this case.  There is a metamorphosis of the gauge field 
into an isotropic optics with refractive index $n = 1/r$.  It has been 
possible to choose a ``host'' Riemannian background which entirely 
incorporates the gauge field.  In that case, there can exist no copies 
(the copy exhibited by Wu and Yang appears in the presence of a source 
current).

In the example above we have taken a solution of the classical field 
equation.  Solutions or not, general field configurations of the form 
\be A^{a}{}_{j} = - \ \epsilon^{a}{}_{j k} \;  \partial^{k} \ln n(x), \ee 
are taken by the dreibeine $h^{a}_{i} = \delta^{a}_{i} \; n $ into 
connections which coincide with the Christoffel symbols of the 
corresponding metric $g_{ij} = \delta_{i j} n^{2} $.  
There is something curious about such cases: a system of coordinates 
exists in which a symmetric connection vanishes.  There is 
consequently a kind of equivalence principle for this type of gauge 
field: by a judicious choice of dreibeine, and then of coordinates, 
the potential (though not the field strength) can be made to vanish.

		\section{Probing into internal space}		%
		
We can use some general characteristics to investigate the ``internal 
geometry'' obtained.  Geodesics, for example, have a strong 
mathematical appeal, and are much used in gravitation to describe 
general qualitative properties of spaces.  It is natural to ask 
whether they have some role here.  In the pure-optics case, as the 
light-ray equation, the geodesic equation does provide an intuitive 
picture of the system.

From the strictly metric--Riemannian point of view, the geodesic 
equation for the general case,

\be
\frac{d  \;  v^{i}}{ds} + {\stackrel{\circ}{\Gamma}}^{i}{}_{j k} \;  v^{j} 
 \; v^{k} = T_{j k}{}^{i}  \;  v^{j} \;  v^{k} \; ,
\ee
can be seen as a kind of force law.  The right-hand side would vanish 
for shortest-length curves.  As it does not, shortest-length curves 
are not self-parallel.  Notice that the affine parameter ``s'' has 
nothing to do with time, and ${\bf v}$ is only a unit vector tangent 
to the curve.  No physical particle is expected to follow such a path.

On the other hand, parallel transport is taken into parallel transport 
by the geometrizing transcription.  A test particle in a gauge field 
is described by (i) its spacetime coordinates and (ii) an ``internal'' 
vector ${\bf I} = \{I_{a} \}$ giving its state in isotopic space.  The 
corresponding dynamic equations (Wong 1970; Drechsler, 1981) are (i) 
the generalized Lorentz force law, which for a unit mass reads

\be   {\frac{d^{2}x^{\mu}}{d\tau^{2}}} = \; I_{a} \; F^{a \mu 
\nu} \; {\frac{dx_{\nu}}{d\tau}} \; , \ee 
and (ii) the so-called charge--precession equation,

\be {\frac{d \; {\bf I}}{d\tau}} \; + {\bf A}_{\mu} \times {\bf I} \; 
{\frac{dx^{\mu}}{d\tau}} \; = \; 0 \; .  \ee The latter says that 
internal motion is a parallel--transport and a precession, as ${\bf 
I}^{2}$ is conserved.  Its transcription, \[ {\frac{d \; I^{i}}{d s}} 
\; + \; \Gamma^{i}{}_{j k} \; I^{j} \; v^{k} = \; \frac{DI^{i}}{Ds} \; 
= \; 0 ,\] says that the transcript of ${\bf I}$ precesses 
parallel--transported by $\Gamma$ along the transcripted curve.  If we 
take for ${\bf I}$ a current ${\bf I} = k {\bf v}$, it gives just the 
geodesic equation.  As to the Lorentz law, it takes the form
 
\be  \frac{d \; v^{i}}{d s} - {\textstyle \frac{1}{2}}\; 
I^{s}{}_{r} \; R^{r}{}_{s}{}^{i j} \; v_{j} = 0 \; .  \ee
This expression, with a velocity-curvature coupling, is more 
akin to the Jacobi than to the geodesic equation. It implies the 
conservation of ${\bf v}^{2}$. Combining  the 
geodesic equation with the charge--precession 
equation, we find 
\[  \frac{d \; (I_{i} \; v^{i})}{d s} = 0. \]  
Thus, ${\bf I}$ 
keeps constant its component along a geodesic.

		\section{Full geometrization}		%

In all we have done previously, the original, ambient space indices 
have been preserved.  Only algebra--related indices have been 
``geometrized''.  This has the advantage of simplicity, as all the 
expressions are written in the initial holonomic basis of ambient 
space.  That space remains what it was, the Euclidean 
3-dimensional flat space, and solutions eventually found will be 
solutions in flat space.  We can now proceed to a complete 
transmutation into curved space, including the ambient space.  We 
identify the two original Euclidean flat spaces and use the dreibeine 
to pass entirely into the new space.  This will lead to more involved 
expressions, as everything will appear written in the anholonomic 
basis defined by the dreibeine.  It will have, however, a double 
merit: we shall be able to speak really of Optics, and new solutions 
will turn up. Indices of both spaces, internal and ambient, become of 
the same kind and can be mixed, as they are in general Relativity. A 
typical example of such mixing is the so-called cyclic identity for 
the Riemann tensor, which comes from (\ref{BianchiFirst}) when 
$T^{a} = 0$. 

To work in an anholonomic basis is, on the other hand, 
a troublesome task.  Thus, after performing the complete 
transposition, it will better to choose a coordinate basis again.  
Coordinates of a 3-space are functions with values in ${\bf E}^{3}$.  
It is particularly interesting to choose the original ambient space 
coordinates as coordinates of the new space because, except for the terms 
involving derivatives, all the above expressions remain formally the 
same.  Thus, (\ref{Ampere}) becomes 
\be \frac{1}{\sqrt{|g|}} \; 
\partial_{j} \left[ \sqrt{|g|} \ R^{r}{}_{s i}{}^{j} \right] + \Gamma 
^{r}_{k j} \; R^{k}{}_{s i}{}^{j} - \Gamma ^{k}_{s j} \; R^{r}{}_{s 
i}{}^{j} = 0 \; .
\label{Amperecurved}
\ee

Notice that this equation is the transcription of the static Amp\`ere 
equation (\ref{AmpLaw}), a particular case of the Yang--Mills 
equations in 4-dimensional spacetime.  Despite its aspect, there is 
{\em a priori} no reason for it to have any special significance by 
itself.  It so happens, however that (\ref{Amperecurved}) is precisely 
the sourceless Yang--Mills equation on the 3-dimensional curved space 
with metric $g$.  This equation is defined (Nowakowski and Trautman, 
1978) on any space as the natural generalization of the flat case: the 
covariant coderivative of the curvature equals to zero (Aldrovandi and 
Pereira, 1995): %
\be *^{-1} d * R + *^{-1} [\Gamma, * R] = 0 \ , 
\label{generalYMeq} \ee %
where $*$ represents the Hodge star operator.  
Some attention must be paid to the signature of the ``host'' metric, 
but in any sourceless case the equation has the component form 
(\ref{Amperecurved}).

And here comes its main interest: it is known (Nowakowski and 
Trautman, 1978; Harnad, Tafel and Shnider, 1980) that the sourceless 
Yang--Mills equation on a symmetric homogeneous space is solved by the 
corresponding canonical connection.  In consequence, any 3-dimensional 
homogeneous symmetric space will provide a solution for 
(\ref{Amperecurved}).  These connections are torsionless, so that we 
come back to pure Optics.  Furthermore, they have constant scalar 
curvature.  Purely Riemannian spaces of constant curvature are not so 
many: they are those hosting the highest possible number of Killing 
vectors.  Given the metric signature and the value of the scalar 
curvature R, there is only one such ``maximally--symmetric'' space 
(Weinberg, 1972), provided torsion is absent.  A negative constant total 
scalar curvature would establish the space as a hyperbolic space.  
Thus, once a complete transmutation is performed, each symmetric 
homogeneous space will provide an Optics which solves the Yang--Mills 
equation.  The simplest 3--dimensional cases are the sphere $S^{3}$ 
and the hyperbolic spaces.  These would be the cases of static 
Yang--Mills equation in Friedmann (respectively closed and open) model 
Universes.

	Consider, to start with, the hypersphere $S^3$ in ${\bf E}^4$, 
	given in Cartesian coordinates $\{ \xi^\mu \}$ by $\sum^4_{\mu = 
	1} (\xi^\mu)^2$ = $(\xi^4)^2 + (\xi^1)^2 + (\xi^2)^2 + (\xi^3)^2 = 
	1$.  We can project it stereographically from the point $\xi^4 = + 
	1$ (its ``north pole'') into the hyperplane ${\bf E}^3$ tangent at 
	the point $\xi^4 = - 1$ (the ``south pole'').  This will provide 
	every point of the hypersphere (except the north pole) with 
	coordinates $x^k = \frac{2 \xi^k}{1 - \xi^4}$ on ${\bf R}^3$.  It 
	is a direct adaptation of the Riemannian metric of $S^3$ on the 
	Euclidean space, with the north pole corresponding to all the 
	points at infinity.  Introducing $r^{2} = 
	\sum^{3}_{i=1}(x^{i})^{2}$ and calculating the line element $ds^2$ 
	= $\sum_{\mu} (\xi^\mu)^2$ in these stereographic coordinates, we 
	obtain the spherical metric $ds^2 = g_{ij}dx^i dx^j$, where $g 
	_{ij} = n^2(x) \delta_{ij}$, with $n = \frac{1}{2} (1 - \xi^4)$ = 
	$\frac{1}{1 + r^2}$.  This case is well known in Geometrical 
	Optics where, with ``n'' the refraction index, it leads to the 
	perfectly-focusing Maxwell fish-eye (Luneburg, 1966).  It does not 
	lead to any bounding in space, as the sphere is taken onto the 
	whole of ${\bf R}^3$.  It is a conformally flat space, as the new 
	metric is at each point proportional to the Euclidean metric.
	
	Take now a hyperbolic space in ${\bf E}^{4}$, given by $(\xi^4)^2 
	- (\xi^1)^2 - (\xi^2)^2 - (\xi^3)^2 = 1$.  It consists of two 
	branches, each one a Lobachevski space.  The points $\xi^4 = + 1$ 
	and $\xi^4 = - 1$ are now the lowest point of the upper branch and 
	the highest point of the lower branch.  The stereographic 
	projection leads now to a metric $g _{ij} = n^2(x) \delta_{ij}$ 
	with the refraction index $n = \frac{1}{1 - r^{2}}$.  In other 
	words, given a hyperbolic metric on ${\bf R}^3$, it is always 
	possible to find a coordinate system $\{ x^{i} \}$ in terms of 
	which the metric is $g _{ij} = n^2(x) \delta_{ij}$, with $n$ as 
	above and $r^2 = \sum^3_{i=1} (x^i)^2$.  Higher-dimensional 
	analogues are the anti-de Sitter spaces, which may exhibit 
	properties analogous to perfect focusing (Hawking and Ellis 1973).
	
	There are two kinds of hyperbolic space, the one-sheeted and the 
	two-sheeted.  Now, it is a well known fact that in the two-sheeted 
	case the above stereographic coordinates divide ${\bf R}^3$ into 
	two parts, one for each branch of the hyperbolic space (Aldrovandi 
	and Pereira, 1995).  One of them is a ball, a Poincar\'e space, 
	the interior (r $< 4$) of a sphere $S^2$ (where r = 4), the other 
	(r $> 4$) its complement in ${\bf R}^3$.  The geodesics are easily 
	computed, and better suited to get some intuition about what 
	happens, showing an ``optics'' with some great differences with 
	respect to Maxwell's fish-eye.  This shows a ``confining'' 
	behavior, which is a global effect of the hyperbolic geometry.  
	Locally, one could be misled by intuition, as neighboring 
	geodesics tend to approach each other in the spherical case, 
	thereby simulating an attraction, and to separate from each other 
	in the hyperbolic case (Arnold, 1978).  The bounding sphere $S^2$ 
	itself is a singular region, corresponding to the infinite regions 
	of both branches.  It plays the role of a ``natural'' bag.  Of 
	course, there is no reason to believe that test particles will 
	follows geodesics, but actually all continuous paths starting 
	inside the region are trapped within it.  There is another point: 
	the metric $g _{ij}$ becomes infinite on the bounding sphere, so 
	does the magnetic field $B$ and, consequently, the energy density 
	in (\ref{Hamiltonian}).  Space is in this way divided into two 
	regions separated by a barrier on which the energy density 
	diverges.  This ``confinement'' remains, of course, of academic 
	interest, because it only occurs when the ambient space is 
	hyperbolic.

Summing up: a Yang-Mills field can, under certain conditions, be 
described as an optical medium on 3-space.  This fact leads to an 
alternative to the usual potential picture as a source of ideas and 
physical intuition.

\section*{Acknowledgments}

The authors are grateful to FAPESP (S\~ao Paulo, Brazil), for 
financial support.  M. Dubois-Violette and R. Kerner are warmly 
thanked for very useful discussions.

	\section*{References}	%

\noindent Aldrovandi, R., and Pereira, J.G., (1995).  {\it An 
Introduction to Geometrical Physics}, World Scientific, Singapore.

\noindent Arnold, V.I., (1978).  {\it Mathematical Methods of 
Classical Mechanics}, Springer, New York (in appendix).

\noindent Calvo, M., (1977).  Phys.Rev.  {\bf D15} 1733.

\noindent Chandrasekhar, S., (1992).  {\it The Mathematical Theory of 
Black Holes}, Oxford University Press.

\noindent Doria, F. A., (1981).  Commun.  Math.  Phys.  {\bf 79} 435.

\noindent Drechsler, W., and Rosenblum, A., (1981).  Phys.Lett.  {\bf 
106B} 81.

\noindent Faddeev, L.D., and Slavnov, A.A., (1978).  {\it Gauge 
Fields.  Introduction to Quantum Theory}, Benjamin/Cummings, Reading, 
Mass.

\noindent Feynman, R.P., (1977).  in {\it Weak and Electromagnetic 
Interactions at High Energy}, Les Houches Summer School of 1976, 
North-Holland, Amsterdam, mainly from page 191 on.

\noindent Freedman, D.Z., and Khuri, R.R., (1994).  Phys.Lett.  {\bf 
B329} 263.

\noindent Freedman, D.Z., Haagensen, P.E., Johnson, K., and Latorre, 
J.I., (1993).  MIT preprint CTP 2238; see also Johnson's contribution 
in {\it QCD - 20 years later}, Aachen, June 1992; and Haagensen's 
lecture at the XIII Particles and Nuclei International Conference, 
Peruggia, Italy, June-July 1993, Barcelona preprint UB-ECM-PF 93/16.

\noindent Greub, W. H., Halperin, S., and Vanstone, R. (1972).  {\it 
Connections, Curvature and Cohomology}, Academic Press, New York.  
Volume II: {\it Lie Groups, Principal Bundles and Characteristic 
Classes}, p.  344.

\noindent Guillemin, V., and Sternberg, S., (1977).  {\it Geometric 
Asymptotics}, AMS, Providence, Rhode Island.

\noindent Harnad, J., Tafel, J., and Shnider, S., (1980).  J.Math.  
Phys.  {\bf 21} 2236.

\noindent Hawking, S.W., and Ellis, G.F.R., (1973).  {\it The Large 
scale structure of space-time}, Cambridge University Press.

\noindent Itzykson, C., and Zuber, J.-B., (1980).  {\it Quantum Field 
Theory}, McGraw-Hill, New York.

\noindent Jackiw, R., (1980).  {\it Rev.Mod.Phys.} {\bf 52} 661.

\noindent Kobayashi, S., and Nomizu, K. (l963).  {\it Foundations of 
Differential Geometry}, 1st vol., Interscience, New York.
 
 \noindent Luneburg, R.K., (1966).  {\it Mathematical Theory of 
 Optics}, University of California Press, Berkeley.
												
\noindent Lunev, F. A., (1992).  Phys.Lett.  {\bf B295} 99.

\noindent Mostow, M. A., (1980).  Commun.  Math.  Phys.  {\bf 78} 137.

\noindent Nowakowski, J., and Trautman, A., (1978).  J.Math.  
Phys.{\bf 19} 1100.

 \noindent Ramond, P., (1981).  {\it Field Theory: A Modern Primer}, 
 Benjamin/Cummings, Reading, Mass.
															
\noindent Roskies, R., (1977).  Phys.Rev.  {\bf D15} 1731.

\noindent Weinberg, S., (1972) {\it Gravitation and Cosmology} J. 
Wiley, New York; mainly chap.  XIII. %
\noindent Wong, S. K., (1970).  Nuovo Cimento {\bf 65A} 689.

\noindent Wu, T.T., and Yang, C.N., (1975).  Phys.Rev.{\bf D12} 3843 
and 3845.

\end{document}